\documentclass[twocolumn,prl,showpacs,superscriptaddress]{revtex4}
\usepackage{graphicx}
\begin{document}
\title{Molecular Mechanism for Nitrogen fixation: first steps}

\author{Johannes Schimpl} \affiliation{Institute for Theoretical
Physics, Clausthal University of Technology, D-38678
Clausthal-Zellerfeld, Germany}

\author{Helena M. Petrilli}

\affiliation{Instituto de F\'{\i}sica, Universidade de S\~{a}o Paulo, Caixa
Postal 66318, 05315-970, S\~{a}o Paulo, SP, Brazil}

\author{Peter E. Bl\"ochl}
\affiliation{Institute for Theoretical Physics, Clausthal University
    of Technology, D-38678 Clausthal-Zellerfeld, Germany}

\begin{abstract}
N$_2$ association to the FeMo-cofactor of nitrogenase, including the
recently identified central N ligand, has been investigated using
first-principles electronic structure calculations.  The oxidation
state of the resting state of the cofactor and its electronic
structure has been identified. A single proton is added to the sulfur
bridges following each electron transfer to the cofactor. During N$_2$
association, the cofactor undergoes large rearrangements resulting in
opening the central Fe-cage of the cofactor.  N$_2$ binds axially
while the bond of the bridging SH group breaks. It is then able to insert
between the two Fe sites in a bridged configuration.
\end{abstract}
\date{\today}
\maketitle 

%==============================================================
%\section{Introduction} 
%==============================================================
Atmospheric nitrogen is the main natural source of nitrogen, which
makes up about 10\,\% of the dry mass of biological matter. Before
nitrogen molecules can be consumed by organisms, they need to be
converted into ammonia, which requires breaking one of the strongest
bonds in nature.  For this purpose, biological nitrogen fixation
employs the enzyme nitrogenase. Nitrogenase consists of two proteins,
the Fe protein and the MoFe protein. The latter contains an
Fe$_7$MoS$_9$ cluster as the proposed active site. This so-called
FeMo-cofactor has been named the most complex bioinorganic cofactor in
nature.  The crystal structure of nitrogenase was unraveled about ten
years ago\cite{kim92,kim92a,geo92}. Despite intense research, however,
the reaction mechanism still remains elusive to date.

A puzzling feature of the FeMo-cofactor was the apparent presence of a
cavity surrounded by four iron sites\cite{kim92,kim92a,geo92}.  Most
previous ab-initio calculations rested on the assumption that the cage
is empty. Recent crystallographic studies\cite{ein02}, however,
identified the presence of a central ligand in the cavity, being a
C,O, or N atom, which sheds new light on the mechanism of biological
nitrogen fixation.

The reaction consumes eight electrons and protons and produces one
sacrificial hydrogen molecule\cite{bur96}. 
\begin{eqnarray*}
\textrm{N}_2+8\textrm{H}^++8\textrm{e}^-\rightarrow 2\textrm{NH}_3+\textrm{H}_2
\end{eqnarray*}
The electron transfer from the Fe protein to the MoFe protein
containing the cofactor is the rate limiting factor for nitrogen
fixation. Electrons are transfered to the cofactor at a rate of about
one to ten per second\cite{fis01}.

A number of reaction mechanisms from nitrogen to ammonia at
the FeMo-cofactor have been proposed.  They can be classified according
to the way N$_2$ binds to the cofactor: (1) Nitrogen binds head on to
one of the six prismatic iron atoms in an ($\eta_1$)
coordination\cite{dan97,dan98,rod99,rod00,rod00a}. (2) Nitrogen forms
a N$_2$ bridge between two octahedrally coordinated Fe atoms after
opening of the cage\cite{tho96,sel99,sel00} and (3) N$_2$ coordinates
to Mo\cite{pic96,gro98,szi01,dur02,dur02a}. (4) Binding of N$_2$ to
the face formed by four iron sites has been ruled out with the
presence of a central ligand\cite{dan03}.

In this paper we investigate the N$_2$ binding modes to the
FeMo-cofactor using state-of-the-art electronic structure
calculations. Inclusion of the central nitrogen ligand dramatically
changes our view of the reaction mechanism: The cage of the FeMo
cluster opens up upon binding to nitrogen, supporting earlier
suggestions that the cluster may undergo major rearrangements during
the enzymatic cycle\cite{tho96}. This indicates that the reaction
mechanism is more complex than previously believed.  Moreover, we find
that N$_2$ binds to the central cage, whereas binding to the Mo site,
a major contender for the role as the reactive site of the cluster, is
thermodynamically unstable.
%==============================================================
%\section{Calculation details.} 
%==============================================================

We performed first-principles electronic structure calculations based
on density functional theory (DFT)\cite{hoh64,koh65} using the PBE
functional\cite{per96}.  We employed the projector-augmented wave
method\cite{blo94} and allowed for a non-collinear description of the
spin distribution. The latter is important to properly account for the
frustrated antiferromagnetically coupled arrangement of high-spin Fe
atoms. The artificial interaction between periodic images of the
cluster in our plane-wave based method has been removed\cite{blo86}.

\begin{figure}
\includegraphics[scale=0.4]{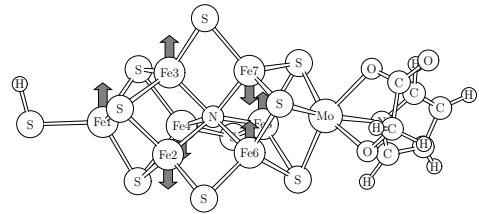}
\caption{Resting state of the FeMo-cofactor.  \label{fig:rs}}
\end{figure}

We considered the complete FeMo-cofactor as shown in
Fig.~\ref{fig:rs}. The central ligand has been chosen to be nitrogen.
The ligands of the FeMo-cofactor have been truncated such that only
single bonds were broken, and the open bonds were saturated by
hydrogen atoms.  Thus we included an imidazole and a glycolate
coordinated to the Mo site to replace the histidine and homocitrate
ligands respectively and an SH group instead of a cysteine group at
the terminal Fe atom of the cofactor.

The interaction of the cofactor with the surrounding protein has
been analyzed using a classical force field, namely the UFF force
field\cite{rap92}.  The protein structure, as obtained from the protein
data bank entry 1QGU\cite{may99}, has been included up to a radius of
10\,\AA{} and held rigid beyond a radius of 9\,\AA{} from FeMoco center
and relaxed inside.

%==============================================================
%\section{Oxidation state} 
%==============================================================
Before exploring the N$_2$ adsorption we need to determine the charge
and protonation state of the cofactor.  Since the driving forces for
protonation and electron transfer are not known a-priori, we derive
them by comparing our theoretical with experimental results. This
implies identifying the charge state of the resting state and to trace
the electron and proton transfer steps until N$_2$ binds.

A reference is provided by the clear $S=3/2$ EPR signal\cite{bur96} of
the resting state. For odd charge states ranging from $-5$\,e to
$+1$\,e, we find that only the charge state $-3$\,e, which is
collinear, can be clearly identified with an S=3/2 spin state.
Charges of $-1$\,e as well as $+1$\,e result in an S=1/2 state, and the
charge state of $-5$\,e has a non-collinear spin distribution with
$S=0.24$. For the noncollinear clusters we determined the spin state
from the absolute value of the integrated spin density. Full
structural relaxation in each charge state has been important for the
determination of the correct ground state as the spin distribution
depends sensitively on the atomic structure.

A charge state $-3$\,e corresponds formally to
Mo$^{4.5+}$Fe$^{2.5+}_5$Fe$^{2+}_2$N$^{3-}$S$^{2-}_9$, with the
ligands contributing a change of $-3$\,e. Six iron atoms form pairs
with a parallel spin alignment.  The pairs are antiferromagnetically
coupled with the neighboring Fe sites as shown in
Fig.\ref{fig:rs}. One Fe atom, located next to the Mo site, remains
unpaired and is antiferromagnetically coupled to all three of its
neighbors. Its spin is oriented in the minority spin direction.

This spin arrangement corresponds to the experimentally observed
distribution of four sites aligned with the main spin direction and
three antiparallel sites, as found in ENDOR\cite{ven86,tru88} and
M\"ossbauer\cite{yoo00} studies. It should be noted that experiments
at higher temperatures may observe an averaged spin structure due to
either electronic or structural fluctuations.

The Fe sites are in a distorted tetrahedral environment formed by
either four S ligands or three S ligands and the central N-ligand,
while the Mo-site is octahedrally coordinated.  The bonding network is
augmented by metal-metal bonds, derived from the Fe e$_g$ and Mo
t$_{2g}$ orbitals. While the metal-metal bonds are fairly delocalized,
the best formal assignment is to attribute mixed valence bonds to the
spin paired Fe atoms and those between Mo and its three nearest
Fe-neighbors.  The mixed-valence bonds are limited to the minority
spin direction of the participating Fe atoms since the majority spin
direction has a filled d-shell. This assignment, derived from an
analysis of the off-site elements of the density-of-states, accounts
for the total charge and spin for the cluster. In addition it explains
the presence of the small magnetic moment of Mo antiparallel to the
main spin direction.

Given the good agreement of our calculated atomic structure for the
unprotonated cofactor with X-ray\cite{ein02} and
EXAFS\cite{har98,chr95} experiments, as opposed to the protonated
cofactor, we conclude that the resting state is indeed unprotonated.

%==============================================================
%\section{Protonation of the cofactor}
%==============================================================

In order to understand N$_2$ adsorption, one needs to know the number
of protons bound to the cofactor.  A reasonable assumption used in our
analysis is that the proton transfer rate is fast compared to the slow
electron transfer\cite{sim84,tho84,low84,tho84a}. This implies that
the protonation state reaches thermal equilibrium before the next
electron is transfered. The protonation state is then determined by
the proton chemical potential reflecting the acidity of the cavity
containing the cofactor. The proton chemical potential, not accessible
in our calculation, will then be calibrated by comparing our findings
with experiment.

In order to determine protonation of the cofactor, we investigated the
protonation energies of all relevant proton acceptor sites for the
singly reduced cofactor.  In accordance with previous calculations
without central ligand\cite{lov02a}, we find that only the bridging
sulfur atoms are protonated. Proton addition to the Fe atoms or the
$\mu_3$ sulfur atoms is less favorable by 0.2\,eV and 0.5\,eV,
respectively. A proton added to the Fe-site converts into a hydride
(H$^{-}$), which can react with a second proton to form a hydrogen
molecule.  

As obtained from collinear calculations, the proton chemical potential
increases by approximately 2.7\,eV per proton added to the sulfur
bridges and decreases by the same amount for each electron added.
This shows that a single proton is added to the cofactor for each
additional electron in a ping-pong like manner. Note that the
dielectric screening by the environment affects the differences of the
calculated protonation energies, but not the qualitative finding of a
ping-pong mechanism. 

The proton transfer is reflected in a structural change of the
cofactor that provides us with a means to relate the protonation state
of the cluster to experiment.  While electron transfer alone does not
change the structure of the cofactor appreciably in our calculations,
the protonation decreases the angle of the sulfur bridges, which in
turn contracts the cluster. EXAFS measurements indicate that the
cluster contracts upon reduction by one electron for
\textit{Azotobacter vinelandii}\cite{chr95}, while no significant
changes have been found for \textit{Klebsiella
pneumoniae}\cite{ead97}.  The fact that proton transfer, as apparent
by the contraction, depends on subtle changes of the protein allows
us to identify the proton chemical potential approximately with the
first protonation energy of the cofactor reduced by one electron.
Knowing the chemical potential we can essentially predict the sequence
of electron and proton transfers.

\begin{figure}[htb]
\center
\includegraphics[scale=0.4]{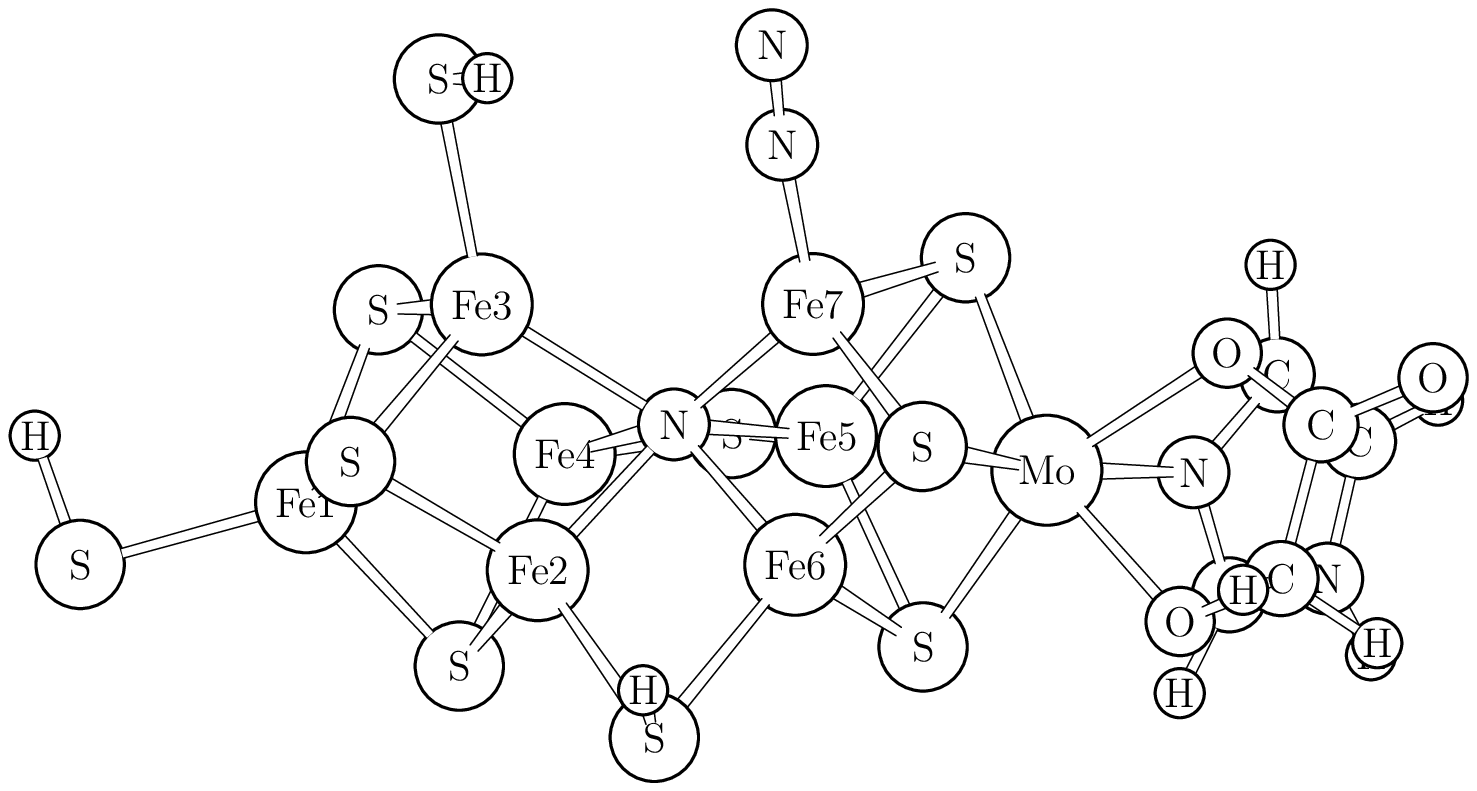}
\includegraphics[scale=0.4]{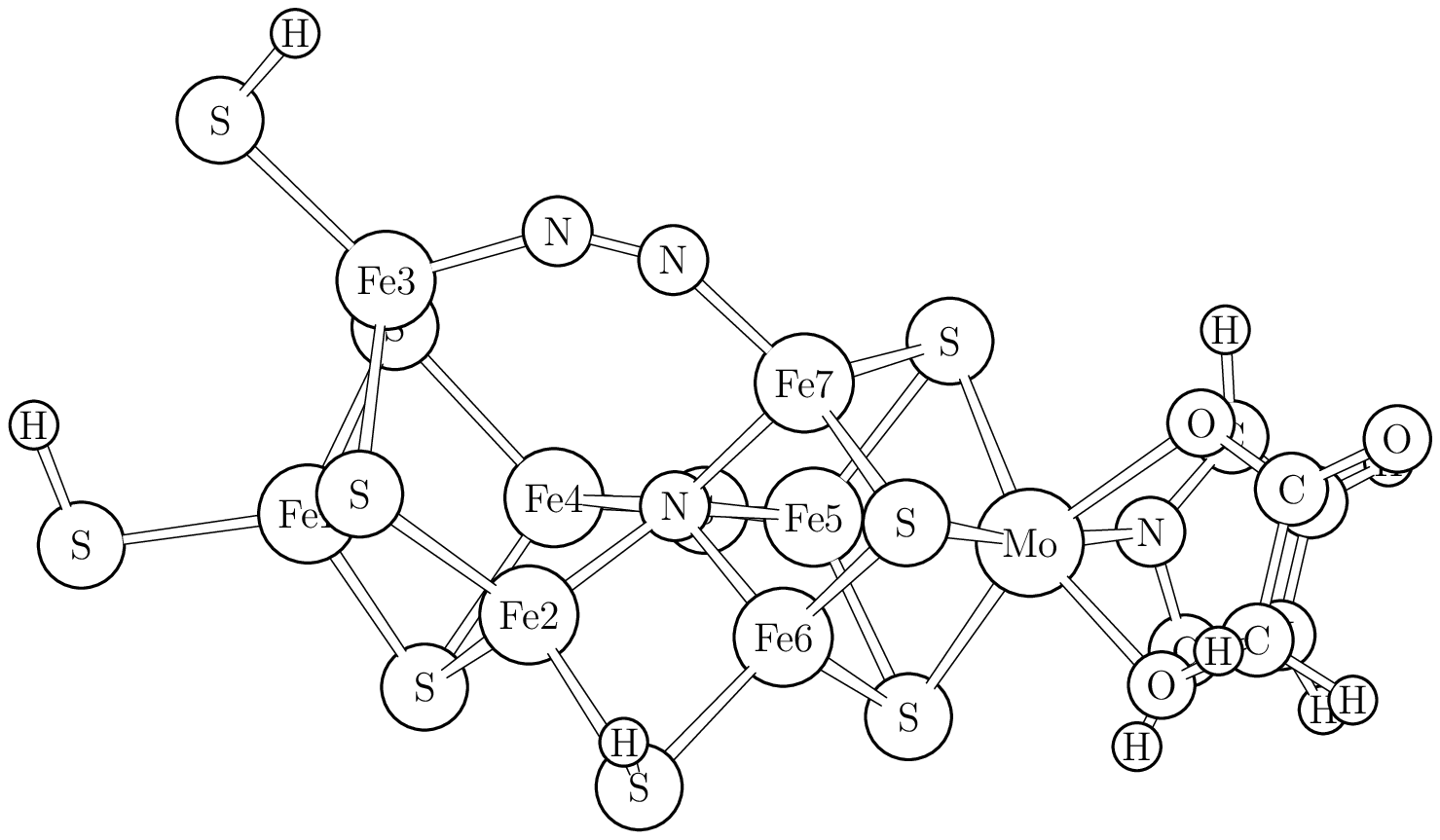}
\includegraphics[scale=0.4]{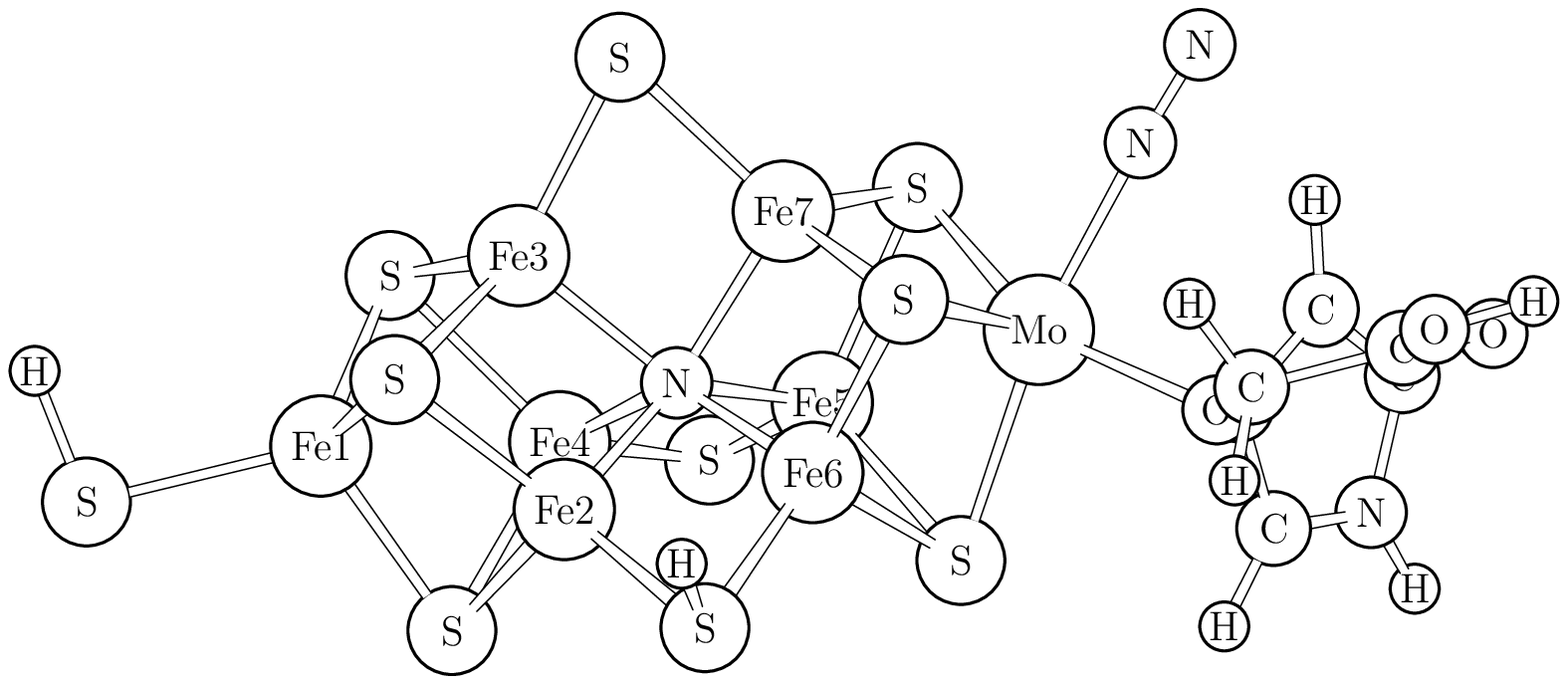}
\caption{Structures of the three N$_2$ binding modes investigated in
  this work: the open axial mode (top), the bridged mode (middle) and
  Mo-coordination (bottom).}
\label{fig:1}
\end{figure}
 
%==============================================================
%\section{N$_2$ binding: axial coordination to Fe.} 
%==============================================================

We have investigated N$_2$ binding after transfer of one and two
electrons with the corresponding number of protons. Our calculations
indicate that N$_2$ binds only in the latter case. Our result that
N$_2$ binds to the doubly protonated cofactor seems to disagree with
Thorneley-Lowe scheme\cite{tho84,low84,tho84a} which predicts that 3-4
electrons reach the MoFe-protein before N$_2$ binding.  However, EPR
measurements during turnover\cite{fis01} indicate that only two of the
three first electrons transfered to the protein actually reach the
cofactor. Thus we need to add one electron (and one proton) before
comparing our results for the cofactor with the Thorneley-Lowe scheme.

On the basis of the doubly reduced cofactor, we investigated several
binding modes of N$_2$: On the faces of the central cage, in the
bridging position between Mo and Fe, at the Mo site and at the
Fe-atoms of the central cage in the axial, equatorial and side-on
orientations.  All of these complexes have been previously discussed
and investigated theoretically for the complex without the central
nitrogen ligand. According to our calculation only the adsorption to
an Fe atom on the central cage is stable.

Upon adsorption of N$_2$ to an Fe-site next to a protonated sulfur
bridge, we find that the sulfur bridge breaks so that the cage
structure of the cofactor is disrupted. Adsorption and cage opening
occur in a concerted mechanism. The resulting structure is shown in
Fig~\ref{fig:1}. The barrier for N$_2$ adsorption is 0.29\,eV, which
can be overcome easily by thermal fluctuations. The binding energy is
0.27\,eV.

For the cofactor without central ligand, Rod et al.\cite{rod99} found
that N$_2$ binds in an axial mode to the same Fe-site, but the cage
structure of the cofactor remained intact in these calculations.  We
find that this result changes radically when the central nitrogen
ligand is included.  Compared to the metastable structure analogous to
that of Rod et al., the cage opening stabilizes N$_2$ binding by
0.55\,eV, indicating that N$_2$ does not bind unless the sulfur bridge
breaks away. The structure with a closed cage is metastable, with a
small barrier of only 0.1\,eV towards the ground state.

With an N-H distance of 2.8\,\AA, the SH group seems to be well
positioned for the first protonation of dinitrogen, which is believed
to have the largest energy barrier in the catalytic cycle.  However,
according to our calculations, this proton transfer is energetically
not favorable.

%==============================================================
%\section{Bridged coordination to Fe.} 
%==============================================================
The axial binding mode is not the only possible configuration for the
N$_2$ complex with the FeMo-cofactor. We find that dinitrogen can tilt
to form a dinitrogen bridge between the two Fe atoms formerly bridged
by a SH group. As dinitrogen binds to the second Fe atom, the bond of
this Fe atom to the central ligand breaks, so that the tetrahedral
coordination of the Fe atom is retained. This bridging configuration
shown in Fig. 2 is energetically more stable by 0.08\,eV than the open
axial mode. The reaction barrier of 0.68\,eV corresponds to a reaction
rate somewhat smaller than the electron transfer rate from the Fe
protein the MoFe protein. Thus both structures, with an axial and
bridged dinitrogen are equally likely intermediates for the N$_2$
fixation cycle.

A similar adsorption mode with N$_2$ bridging two Fe sites has been
proposed earlier by Sellmann et al.\cite{sel99} Sellmann's model
differs from our bridged complex in that the Fe sites are octahedrally
coordinated, instead of the tetrahedral coordination in our
cluster. The different coordination reflects in a major change of the
electronic structure: The octahedral complex results in low-spin Fe
atoms while the tetrahedral coordination results in high-spin
Fe-atoms, which have different chemical behavior. The chemical
analogy to octahedral low-spin complexes\cite{sel96} has been one of the
main reasons for Sellmann's proposal.

The additional ligands in Sellmann's model are water molecules and the
nitrogen atoms from two amino acids of the protein, glutamine
Gln$\alpha$191 and histidine His$\alpha$195\cite{protein}. We have
investigated the model proposed by Sellmann by modeling the nitrogen
ligands with ammonia molecules.  We find this structure at least
metastable in the absence of the central nitrogen ligand. Addition of
the central ligand, however, results in the spontaneous desorption of
the three water ligands from the two bridged Fe sites. The ammonia
ligands remain bound to the Fe sites, so that the latter assume a
pentacoordinate coordination with high-spin iron atoms.

%==============================================================
%\section{Embedding in the protein environment}
%==============================================================
An important question is if the protein environment is able to
accommodate the expansion of the cage after N$_2$ binding. Therefore
we embedded the rigid FeMo-cofactor as obtained from our calculations
into the protein simulated with classical force fields.

The cofactor with N$_2$ adsorbed at sites Fe3 and Fe7 can easily be
accommodated both in the open-cage and in the bridged configuration.
There is also sufficient space to accommodate N$_2$ bound to Fe6, even
though the embedding energy is 1\,eV higher than that for N$_2$ bound
to Fe3 or Fe7.  The presence of a nearby imidazole ring of the protein
(His$\alpha$195)\cite{protein} prevents the transition to the bridged
configuration as well as adsorption to site Fe2.  Binding at sites Fe4
and Fe5 can be excluded, because either the dinitrogen or the SH group
collides with the protein backbone in close proximity of site Fe4.

We conclude that the adsorption complexes can be accommodated in the
central cage. The most likely adsorption sites are Fe3 and Fe7. Given
the uncertainties of the force field calculations we do not
explicitely exclude adsorption to Fe6.

%==============================================================
%\section{N$_2$ binding the Mo-site}
%==============================================================

Coordination of N$_2$ to Mo has been discussed in great detail in the
literature\cite{pic96,gro98,szi01,dur02,dur02a}.
%The
%analogy with the known N$_2$ chemistry in Mo complexes\cite{??} has
%been one reason to attribute a special role to the Mo atom.  
While the molybdenum atom in the cofactor stands out, it is not
essential.  There are other nitrogenases, where the Mo atom of the
cofactor is replaced by V or Fe.  While their efficiency is reduced
their mere existence indicates that the Mo-site is not essential.

The Mo atom is octahedrally coordinated to three sulfur sites of the
cofactor, to two oxygen atoms of homocitrate and to the nitrogen atom
of a histidine. N$_2$ association on the Mo atom is initiated by a
proton transfer to the carboxyl group of homocitrate.  After
protonation, the Mo-O bond becomes very labile. Nevertheless, N$_2$
binding to the vacant coordination site at Mo is unstable by
0.31-0.34\,eV irrespective of the protonation state of the carboxyl
group of homocitrate.  While hydrophobic forces of the environment,
not considered in this work, may increase the affinity to N$_2$, the
presence of more stable binding modes at the Fe sites provides strong
evidence that the mechanism does not proceed at the Mo site.

Our finding that binding of N$_2$ at the Mo site is unfavorable
differs from previous conclusions\cite{szi01,dur02} derived from
smaller model systems. This difference has been traced to the small
cluster size, that is 1-2 metal sites, used in those calculations.

%==============================================================
%\section{Conclusion} 
%==============================================================

In this work, we analyzed the N$_2$ adsorption at the FeMo-cofactor of
nitrogenase containing the recently detected central nitrogen
ligand. Special attention has been given to oxidation states and the
sequence of electron and proton transfer steps. We find that N$_2$
binding results in a disruption of the cage structure of the
FeMo-cofactor.  In contrast to the obvious assumption that the central
ligand adds rigidity to the cofactor, the additional nitrogen atom
offers a variable number of bonds to its Fe neighbors and thus
adds flexibility to the structure.

We acknowledge support by the HLRN for granting access to their IBM
pSeries 690 Supercomputers.

%\bibliographystyle{apsrev}
%\bibliography{make}

\end{document}